\documentclass{article}
\usepackage[utf8]{inputenc}
\usepackage{fullpage}
\usepackage{cite}
\usepackage{tabularx}
\usepackage{amssymb}
\usepackage{lineno,hyperref}
\usepackage{amsmath}
\usepackage{cases}
\usepackage{booktabs}
\usepackage{multirow}
\usepackage[table,xcdraw]{xcolor}
\usepackage{algorithm}
\usepackage{algpseudocode}
\usepackage{graphicx}
\usepackage{hhline}
\usepackage{subfig}
\usepackage{caption}
\usepackage{lipsum}

\begin{document}

\title{Recovery and Analysis of Architecture Descriptions   using Centrality Measures}

\author{Sanjay Thakare$^1$, Arvind W Kiwelekar$^2$\\ 
 Babasaheb Ambedkar Technological University,\\Maharashtra, India.\\  1:mail2sbt@gmail.com, 2:awk@dbatu.ac.in}   

\maketitle

\begin{abstract}
The necessity of an explicit architecture description has been continuously emphasized to communicate the system functionality and for system maintenance activities. This paper presents an approach to extract architecture descriptions using the {\em centrality measures} from the theory of Social Network Analysis. 
The architecture recovery approach presented in this paper works in two phases. The first phase aims to calculate centrality measures for each program element in the system.  The second phase assumes that the system has been designed around the layered architecture style and assigns layers to each of the program element. Two techniques to assign program elements are presented. The first technique of layer assignment uses a set of pre-defined rules, while the second technique learns the rules of assignment from a pre-labelled data set.   The paper presents the evaluation of both approaches.  
\end{abstract}


\section{Introduction}

The value of explicit software architecture has been increasingly recognized for software maintenance, and evolution activities \cite{link2019value}. Especially   architecture descriptions relating coarsely granular programming elements   found  as a useful   tool to effectively communicate system functionality and architectural decisions \cite{pacheco2018designing,venters2018software}. These descriptions also support dependency analysis which drives the task of software modernization \cite{escobar2016towards}. Despite the numerous benefits,  a legacy or open-source software system often lacks such kind of architecture descriptions. Moreover when such architecture descriptions are available, these are not aligned with the latest version of system implementation  \cite{shahbazian2018recovering}.  

In such situations, a light-weight architecture recovery approach which approximately represents the true architecture of a system may be more convenient over the sophisticated techniques of architecture recovery. Such a light-weight approach  shall quickly extract relevant information necessary to build architecture descriptions  so that it can provide much-needed assistance to software architects dealing with re-engineering and modernization of existing systems, thus increasing their productivity. 

Intending to design a light-weight approach, this paper presents an architecture recovery approach based on {\em centrality measures} from the theory of Social Network Analysis \cite{papacharissi2009virtual,albert2002statistical}. Three observations drove the rationale behind using centrality measures  for architecture extraction:
(i) Most of these measures provide a highly intuitive and computationally simple way to analyze interactions when a graph represents the structure of a system.
(ii) These measures quantify the structure of a system at multiple levels, i.e.,  at a particular node level, in relation to other nodes in the graph, and at a group of nodes or communities.
(iii) These measures support the development of data-driven approaches to architecture recovery.

The centrality measures-based approach presented in this paper recovers architecture descriptions in two phases. In the first phase, a centrality score is assigned to each program element.  We assume that the system functionality is decomposed among  multiple layers, and  so in the second phase, a layer is assigned to each program element. 

The paper primarily contributes to existing knowledge-base of architecture recovery domain in the following ways. (1) The paper demonstrates the use of centrality measures in recovering high-level architecture descriptions.  (2) The paper describes a data-driven approach to architecture recovery using supervised classification algorithms.
(3) The paper presents an evaluation of supervised classification algorithms in extracting architectural descriptions. 

Rest of the paper is organized as follows: The centrality measures used in the paper are defined in Section II.  Section III describes the central element of the approach. The algorithmic and data-driven approaches to the problem of layer assignment are explained in Section IV.  The results and evaluation of the approach are presented in Section V.  The section VI puts our approach in the context of existing approaches by discussing its features in relation with them. Finally, the paper concludes in Section VII.

\begin{figure*}[t]
	\centering
	\caption{An Example of Class Dependencies with  and their  centrality scores.}
	\label{f1}
\begin{tabular}{c}	
\includegraphics[scale=0.11]{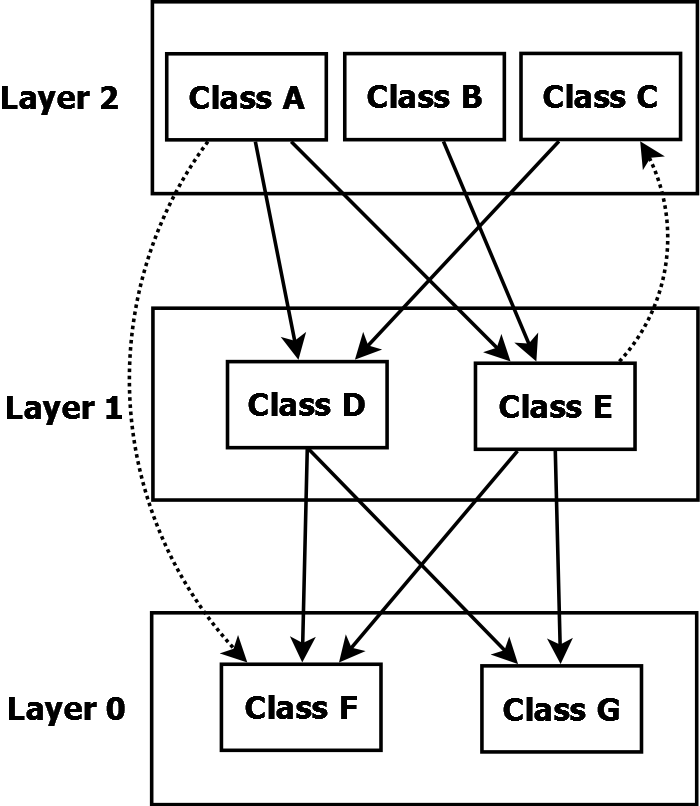}
\end{tabular}	
	\begin{tabular}{|c|c|c|c|c|c|c|}
		\hline
	  
	PID	& ind & outd & deg & bet & clos & eig  \\ \hline
		A & 0 & 3 & 3 & 0 & 0.71 & 0   \\ \hline
		B & 0 & 1 & 1 & 0 & 0.5 & 0   \\ \hline
		C & 1 & 1 & 2 & 2 & 0.6 & 0.055   \\ \hline
		D & 2 & 2 & 4 & 2.5 & 1 & 0.27   \\ \hline
		E & 2 & 3 & 5 & 5.5 & 0.8 & 0.0055  \\ \hline
		F & 3 & 0 & 3 & 0 & 0 & 1  \\ \hline
		G & 2 & 0 & 2 & 0 & 0 & 0.99 \\ \hline
		L2 & 1 & 5 & 6 & 0 & 1 & 0.5   \\ \hline
		L1 & 4 & 5 & 9 &  1 & 1 & 0.5   \\ \hline
		L0 & 5 & 0 & 5 & 0 & 0 & 1  \\ \hline
	\end{tabular}
\end{figure*}

\section{Social Network Analysis Measures} \label{measures}

 The theory of Social Network Analysis (SNA) provides a generic framework to analyze the structure of complex systems. This framework includes a rich set of measures, models, and methods to extract the patterns of interactions among systems' elements. A complex system is expressed as a network of nodes and edges to support the analysis of systems from diverse application domains.

 Examples of complex systems that have been analyzed with the help of SNA include communities on social media platforms\cite{papacharissi2009virtual}, and neural systems\cite{albert2002statistical}. The techniques from SNA have been applied   
 to control the influence of the disease\cite{watts1999networks} to understand  biological systems\cite{silva2015methodology}, to investigate the  protein interactions \cite{AMITAI20041135},  and  to examine animal behavior\cite{wey2008social}. 
  
 These diverse applications of SNA show that  a complex system exhibits certain graph-theoretic common properties  such as centrality, scale-free, small world, community structure, and power-law degree distribution \cite{newman2002random,newman2001random,newman2003structure,albert2002statistical,borgatti2005centrality,freeman1979centrality}.  Some of these commonly observed SNA measures relevant to our study are described below.

The theory of Social Network Analysis (SNA) provides a range of measures with varying levels. Some are applied at the node level where others applied at the network level. The node-level measures are fine-grain measures that are calculated from the nodes which are directly connected to a  given node.    The {\em Centrality measures}\cite{singh2020centrality} are the examples of node-level measures that quantify the importance of an individual or a node in the network. A central node is an influential node having significant potential to communicate and access the information.  There  exists different {\em centrality measures} and they are derived from the connections to a node, position of a node in the network, distance of a node from others, and relative importance of nodes.

\subsection{Degree centrality}	This measure determines the central node based on the connections to the individual node. A node with a higher degree in the network is considered  as the most influential one.  In a directed graph, two different centrality measures exist {\em in-degree} and {\em out-degree} based on the number of incoming and outgoing edges respectively. The degree centrality of a node $v$ is equal to the number of its connections normalized to the maximum possible degree of the node.
\begin{equation}
C_{D}(v) = deg(v)     
\end{equation}
\begin{equation}
NC_{D}= \frac{C_{D}(v)}{n - 1}=\frac{deg(v)}{n - 1}    
\end{equation}
 
\subsection{Closeness centrality}
This measure aims to identify an influential node in terms of  faster and wider spread of information in the network. The influential nodes are characterized by a smaller inter-node distance which signify the faster transfer of information. The closeness centrality is derived from the average distance from a node to all the connected nodes at different depths. However, the distance between the disconnected components of the network is infinite and  hence it is excluded. For the central node, the average distance would be small and is calculated as the inverse of the sum of the distance to all other nodes. The normalized closeness($NC_{C}$) is in the range from 0 to 1 where 0 represents an isolated node and 1 indicates a strongly connected node.
\begin{equation}
C_{C}(v) = \sum ( \frac {1} {d_{vw} }  )    
\end{equation}
\begin{equation}
NC_{C} = \sum ( \frac {n-1} {d_{vw} } )    
\end{equation}

\subsection{Betweenness centrality}
This measure aims to identify those central nodes which are 
responsible for connecting two or more components of the network. Removal of  such a central node would  mean a disconnection of the complete network. Hence,  these nodes act as a bridge  to pass the information \cite{borgatti2005centrality, white1994betweenness}.  Betweenness centrality  is defined as the number of shortest paths passing through a node.
\begin{equation}
C_{B}(v) = \sum_{s \neq v \neq t} \frac{\sigma_{st}(v)}{\sigma_{st}}
\end{equation}
where, $\sigma_{st}$ is the total number of shortest paths from a node $s$ to $t$ and $\sigma_{st}(v)$ is the number of paths that pass through $v$. The relative betweenness centrality of any node in the graph with respect to the maximum centrality of the node is calculated from $C_{B}(v)$. 
\begin{equation}
C_{B}^{'}(v) = \frac {2  C_{B}(v)} { n ^{2} - 3n + 2}    
\end{equation}
	
\subsection{Eigenvector centrality}
	The Eigenvector centrality is a relative centrality measure, unlike the previous three measures that are absolute one. The Eigenvector centrality calculation depends on the largest real Eigenvalue present in the symmetric adjacency matrix. The centrality of a node $v$ is proportional to the sum of the centralities of the nodes connected to it \cite{bonacich2007some,borgatti2005centrality}.
	
	\begin{equation}
	    \lambda v_i =\sum_{j=1}^{n}{a_{ij}v_j}    
	\end{equation}
	
	\par In general, it requires the solution of the equation $Av = \lambda v$  where $A$ is an adjacency matrix. 

 Figure \ref{f1} shows the centrality scores of various programming elements calculated by considering the dependencies as shown in the figure. Here, it is to be noted that centrality scores can be calculated at different granularity levels i.e., at the object, method, class, package or at a logical layer.  In the figure, centrality scores are calculated at class and layer level.  Here, we have considered a layer as logical encapsulation unit loosely holding multiple classes.

\section{Approach}

The broad objective of the approach is to extract high-level architecture descriptions from the implementation artefacts so that the analysis specific to an architecture style can be performed. In this paper, we  demonstrate the approach with the help of implementation artefacts available in a Java-based system such as Java and JAR files. However, the method can be extended to other language-specific artefacts. Similarly, the approach assumes that a system under study is implemented around the Layered architecture style and demonstrates analyses specific to the Layered architecture style.  
\begin{figure}[h]
	\centering
	\includegraphics[scale=0.3]{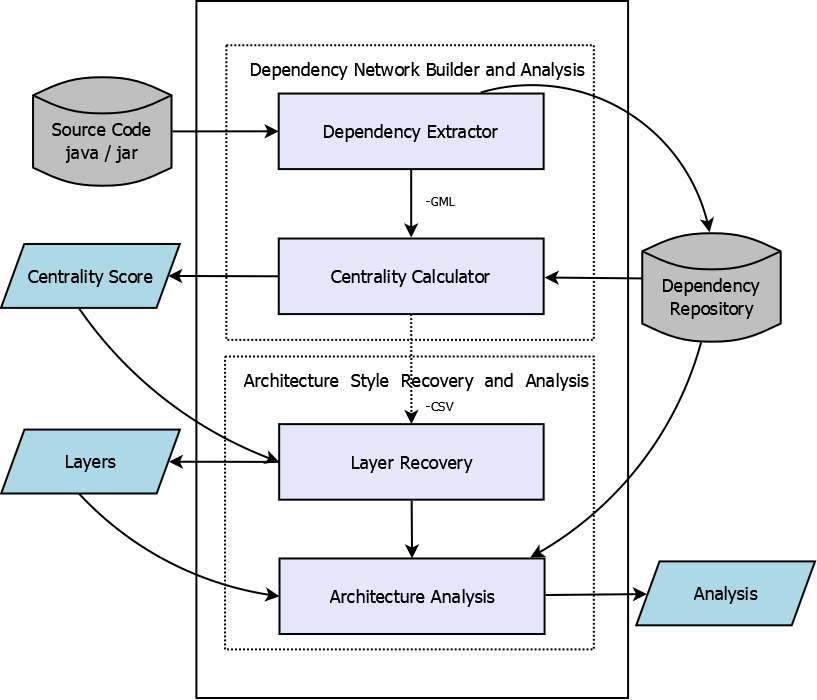}
	\caption{Block diagram of a tool implement in Java to discover layered architecture using centrality.}
	\label{fig:architecture}
\end{figure}

As shown in   Figure \ref{fig:architecture}, the approach consists of  following two phases. 

\begin{enumerate}
	\item \textbf{Dependency Network Builder and Analysis [Phase 1]:} The purpose of this phase is to retrieve the dependencies present in the implementation artefacts. For a Java-based system, this phase takes Java or Jar files as an input and generates a dependency network. The programming elements such as {\em Classes, Interfaces}, and {\em Packages} are the nodes of the network and the Java relationships such as $extends$, $implements$, and $imports$ are the edges in the dependency network.  The output of this stage is represented as a graph in Graph Modeling Language (GML) notations.
	In the second stage, a centrality score to each node is assigned.  The centrality score includes the different measures described in Section \ref{measures}, and they are calculated at the Class and Interface levels.  The output of this stage is a data file in the CSV format describing the centrality score assigned to each program element.
	
	\item {\bf Architecture Style Recovery and Analysis [Phase 2]} The purpose of this phase is to perform architecture style-specific activities.  In this paper, the activities of this phase are illustrated by assuming Layered architecture style. For the Layered architecture style, we define a sub-activity called layer assignment.   The layer assignment activity aims to assign the most appropriate layer to a program element.
	
	Additional  style-specific analyses such as analysis of layer violations,  and performance modeling can be supported once the programming elements are assigned to appropriate layers. 
\end{enumerate}
 The  {\em Phase 1} activities which include building a dependency network and calculating centrality scores are straightforward to realize when the tools such as $JDependency$ are available.  The {\em Phase 2} activities that are performing style-specific analyses can be realized in multiple ways.  Two such techniques to realize architecture style-specific analyses are described in the following section.
 
\begin{table}[]
	\centering
	\begin{tabular}{|l|c|c|c|}
		\hline 
 		\textbf{Centrality} & \textbf{Upper} & \textbf{Middle} & \textbf{Lower} \\ 
		\hline 
		In-degree & low & - & high \\ 
		\hline 
		Out-degree & high & - & low \\ 
		\hline 
		Betweenness & low & high & low \\ 
		\hline 
		Closeness & high & high & low \\ 
		\hline 
		Eigenvector & low & low & high \\ 
		\hline 
	\end{tabular}\\
	\caption{ Relative Significance  of centrality measures with respect to layers}
	\label{relation}
\end{table}

\begin{table}[]
    \centering
    
\begin{tabular}{|p{0.75in}|p{2.25in}|}
\hline
  $\delta_{il}$ and  $\delta_{iu}$  & Lower and upper bound for in-degree centrality values.  \\ \hline 
  $\delta_{ol}$ and  $\delta_{ou}$  & Lower and upper bound for out-degree centrality values.  \\    \hline   $\delta_{b}$   &  Critical Value for between-ness centrality  \\    \hline 
    $\delta_{c}$   &  Critical Value for closeness centrality  \\    \hline 
      $\delta_{e}$   &  Critical Value for eigen-value centrality  \\    \hline 
\end{tabular}

    \caption{Configuration Parameters}
    \label{cp}
\end{table}

\section{Layer Assignment}

The objective of the layer assignment stage is to identify the most appropriate layer based on the centrality measures. We assume a {\em three-layers} based decomposition. 
Here, we use the term {\em layer} in the loose sense that a {\em layer} is a logical coarse-grained unit of encapsulating program elements and not in the strict sense as that used for {\em Layer architecture style} \cite{buschmann2007pattern}. The decision to decompose all the system responsibilities into three layers is based on the observation that functionalities for the majority of the applications can be cleanly decomposed into three coarse-grained layers. For example, many applications typically use architectural styles such as Model-View-Controller (MVC), Presentation-Abstraction-Control (PAC),\cite{buschmann2007pattern}  and a 3-tier style, i.e. Presentation, Business Logic, and Data Storage.

\begin{algorithm} \caption{\textbf{: primaryLabel(inDegree, outDegree, n)}
		\newline
		\textbf{Input:}  inDegree[1:n],outDegree[1:n]: Vector, 
		n:Integer\newline 
		\textbf{Output:} inPartition[1:n], outPartition[1:n]  Vector  }
	\label{algo_CentLayer1}
	\begin{algorithmic}[1]
	  	
		\State Initialize $\delta_{iu}$, $\delta_{il}$, $\delta_{ou}$ and $\delta_{ol}$ 
		\For{\textit{node} in 1 to n}
		\If{$in(node) = 0 $ and $out(node) = 0$}
			\State $inPartition[node] \leftarrow lower $   \State $outPartition[node] \leftarrow lower $ 
5

		\Else	
			\If{$in(node) > \delta_{il} $}
				\State $inPartition[node] \leftarrow lower $ 
			\Else
				\If{$in(node) < \delta_{iu} $}
					\State $inPartition[node] \leftarrow upper $ 	
				\Else
					\State $inPartition[node] \leftarrow middle $ 	
				\EndIf
			\EndIf
			\If{$out(node) > \delta_{ou} $}
				\State $outPartition[node] \leftarrow upper $ 
			\Else
				\If{$out(node) < \delta_{ol} $}
					\State $outPartition[node] \leftarrow lower $ 	
				\Else
					\State $outPartition[node] \leftarrow middle $ 	
				\EndIf
			\EndIf		
		\EndIf
		\EndFor
	\end{algorithmic}
\end{algorithm}

Two different techniques are developed to assign layers to program elements based on centrality measures. The first techniques uses a set of pre-defined rules. The second technique automatically learns the assignment rules from the pre-labelled layer-assignments using a supervisory classification algorithm.
\subsection{Rule-Driven Layer Assignment}

Dependencies among the program elements are used to identify logical units of decomposition. These dependencies are quantified in-terms of centrality measures described in Section \ref{measures}. The  measure of {\em degree centrality}  from Section II-A is further divided as {\em in-degree ($inDeg$)} and {\em out-degree   ($outDeg$)} measures which count the  number of incoming and outgoing edges of a node.
Total five measures of centrality are used.  A set of configuration parameters, as shown in Table \ref{cp}, are defined. These parameters provide flexibility while mapping program elements to a specific layer.  

Five accessor functions namely $in$, $out$, $between$, $closeness$ and $eigen$ are defined to get the values of in-degree centrality, out-degree centrality, betweenness centrality, closeness centrality and eigenvector centrality  associated to a specific node. These functions are used to assign a program element to a specific layer from three layers, i.e. {\em upper, middle and lower}.  
Table \ref{relation} describes the relative significance of various centrality measure regarding upper, middle and lower layers.

The     {\bf Algorithm}  \ref{algo_CentLayer1}  is operated on a  dependency-network in which nodes represent program element and edges dependencies.
The objective of this algorithm is to partition the node space representing into three segments corresponding to lower, middle and upper layers. The algorithm calculates two different partitions. The first partition i.e. $inPartition$ is calculated using the  {\em in-degree} centrality measure while the second partition is calculated using the {\em out-degree} centrality measure.   
\begin{algorithm} \caption{\textbf{refineLabel(inParticion, outPartition, n)}\newline
		\textbf{Input:}  inPartition[1:n], outPartition[1:n]: Vector \newline   
		n: Integer \newline 
		\textbf{Output:}  nodeLabels[1:n]: Vector}
	\label{algo_CentLayer2}
	\begin{algorithmic}[1]
	\State Initialize $\delta_{b}$, $\delta_{c}$ and $\delta_{e}$ 
		\For{\textit{node} in 1 to n}
		
			\If{$inPartition[node] = outPartition[node] $}
				\State $nodeLabels[node] \leftarrow outPartition[node] $ 	
			\Else
				\State $nodeLabels[node] \leftarrow \newline 
					 upDown(inPartition[node],   outPatition[node]) $ \hfill	
			\EndIf

		\EndFor
		
	\end{algorithmic}
\end{algorithm}	

After the execution of {\bf Algorithm} \ref{algo_CentLayer1}  each node is labeled with two labels corresponding to layers. The various combination of  labels include {\em (lower, lower), (middle, middle), (top, top), (middle, top),} and  {\em (middle, lower)}.  Out of these six labelling, the labels {\em (middle, top),} and {\em (middle, lower)} are conflicting  because  two different labels are assigned to a node. This conflict needs to be resolved.

The  {\bf Algorithm}  \ref{algo_CentLayer2} resolves the conflicting labels and assigns the unique label to each node. The conflicting labels are resolved by using  the rules described in the decision Table  \ref{dt}.  The function $upDown$ called in the {\bf Algorithm}  \ref{algo_CentLayer2} uses these rules.  The rules  in Table \ref{dt} resolve the conflicting assignments using the  centrality measures of {\em closeness, between-ness}, and {\em Eigen vector}, while the primary layer assignment is done with {\em in-degree} and {\em out-degree} centrality measures. When   {\bf Algorithm}  \ref{algo_CentLayer2} is executed, some of the nodes from the middle layer bubble up to the upper layer, and some nodes fall to the lower level. Some nodes remain at the middle layer. The vector $nodeLabels$ holds the unique labelling of each node in the dependency network after resolving all conflicts.

\begin{table}[t]
	\centering
	\caption{Decision Table used to Refine Layering}
	\label{dt}
	\begin{tabular}{|c|c|c|p{10.0cm}|}
		\hline
		\multicolumn{1}{|l|}{Layer} & Measure & \multicolumn{1}{l|}{Significance} & \multicolumn{1}{c|}{Rationale}                                                                                                 \\ \hline
		\multirow{3}{*}{upper}        & in   & 0                          & Classes with in-degree value equal to 0 are placed in the upper layer.          \\ \cline{2-4} 
		& out  & high                          & Classes with high out-degree are placed in the top layer because they   use services from layers beneath them.                               \\ \cline{2-4} 
		& closeness   & high                          & Classes with high closeness value are placed in the upper layer because of large average distance from top layer to bottom layer. \\ \hline
		middle                    & between & high                          & Classes with high betweenness value are placed in the middle layer as they fall on the path from top layer to bottom layer.  \\ \hline
		\multirow{4}{*}{lower}     & in   & high                          & Classes with high in-degree value are placed in the bottom layer because they are highly used.                               \\ \cline{2-4} 
		& out  & 0                          & Classes with out-degree value equal to zero are placed to bottom layer because they only provide services.                           \\ \cline{2-4} 
		& eigen & 1                          & Classes with eigen value equal to 1 are placed to bottom layer because they are highly reused.                                \\ \cline{2-4} 
		& in         & -                         & Classes with in-degree and out-degree values are equal to 0 are placed to bottom layer, because they are isolated classes.    \\ \hline
	\end{tabular}
\end{table}

\subsection{Supervised Classification based Layered Assignment}

The {\em configuration parameters} need to be suitably initialized for the correct functioning of the {\em algorithmic-centric} approach discussed in the previous section.   The system architect responsible for architecture recovery needs  to fine tune the parameters to get layering at the desired level of abstraction. To overcome this drawback a {\em data-driven} approach is developed to assign labels to the programming elements.

\begin{table*}[t]
	\centering
	\caption{Sample observations from the Datasets used  for Supervised Learning}
	\label{ds}
	\begin{tabular}{|p{0.25in}|p{1.5in}|p{0.5in}|p{0.6in}|p{0.5in}|p{0.6in}|p{0.5in}|p{0.3in}| }
		\hline 
		 Id	& Label &	In-Degree&	Out-Degree&	Closeness &	Between- ness &	Eigen- vector &	Layer \\ \hline
\multicolumn{8}{|p{5.25in}|}{\centering \text{HealthWatcher}} \\ \hline 		 
1 &	ComplaintRecord	 &1&	10&	1.714&	19&	0.0056&	2 \\ \hline 
2 &	ObjectAlreadyInserted Exception&	37&	0&	0&	0&	0.347&	1 \\ \hline
3 &	ObjectNotFound Exception&	53&	0&	0&	0&	0.943&1 \\ \hline
4&	ObjectNotValidException	&41&	0&	0&	0&	0.883&	1 \\ \hline
5&	RepositoryException&	60&	0&	0&	0&	1&	1 \\ \hline
\multicolumn{8}{|p{5.25in}|}{\centering \text{ConStore}} \\ \hline

1&	Cache&	2&	1&	1&	0&	0.0162&	2  \\ \hline
2&	CacheObject	&4&	0&	0&	0&	0.053&	2  \\ \hline
3&	LRUCache&	0&	2&	1&	0&	0&	2  \\ \hline
4&	MRUCache&	1&	2&	1&	7&	0.0246&	2  \\ \hline
5&	ItemQuery&	1&	20&	0.412&	47.166&	0.0388&	2  \\ \hline
	\end{tabular}
\end{table*}

In the data-driven approach, the problem of layered assignment is modeled as a  multi-class classification problem with three labels i.e. {\em lower (1), middle (2) and upper (3)} with numerically encoded as 1,2, and 3 respectively.  The classification model is trained on the labeled data-set.  The data set, as shown in  Table \ref{ds}, includes program element identifiers, values of all the centrality measures   and layering labels as  specified by the system architect responsible for architecture recovery.   The layering labels can be used from the previous version of the system under study or the labels guessed  by system architect to explore  different alternatives for system decomposition. 

We implement three supervised classification algorithm namely  K-Nearest Neighbour, and Support Vector Machine and Decision Tree. These are the machine learning algorithms particularly used  for multi-class classification problems.  A detailed comparison of these various algorithms can be found in \cite{hassan2018comparison} Python's Scikit-Learn \cite{hao2019machine} library is used to  develop classification model based on these  algorithms.   Table \ref{ds} shows the   format of the  sample dataset used to train the classification  models. The developed models are evaluated against the classification metrics such as accuracy, precision, recall, and F1-Score.
.\begin{figure}[h]
	\centering
	\includegraphics[ scale=0.3]{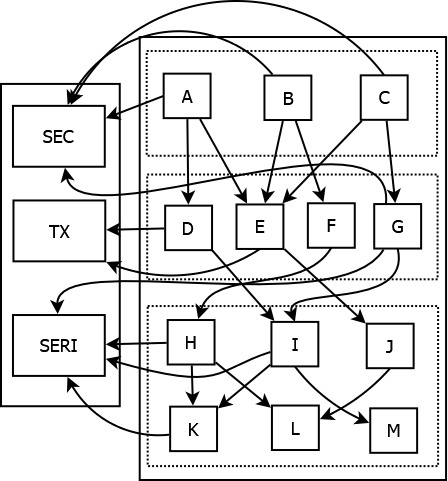}
	\caption{An Architecture of a System Designed to Test the approach}
	\label{tc}
\end{figure}
\begin{table*}
\caption{Accuracy and Confusion Matrix for Data-Driven and Algorithmic Approach} \label{acc}
\begin{center}
\begin{tabular}{|p{0.3in}|p{0.3in}|p{0.3in}|p{0.3in}||p{0.3in}|p{0.3in}|p{0.3in}||p{0.3in}|p{0.3in}|p{0.3in}||p{0.3in}|p{0.3in}|p{0.3in}|}
\hline 
\multicolumn{13}{|p{3.9in}|}{\centering {\bf Constore (Size: 66 classes or interfaces) Confusion Matrix} } \\ \hline 
\multicolumn{4}{|p{1.2in}||}{\centering SVM} &
\multicolumn{3}{|p{0.9in}||}{\centering Decision Tree} &
\multicolumn{3}{|p{0.9in}||}{\centering KNN classifier} &
\multicolumn{3}{|p{0.9in}|}{\centering Rule based} 
\\ \hline 
Layer & lower& middle & upper  & lower& middle & upper  &lower& middle & upper  & lower& middle & upper  \\ \hline 
lower & 43 &0 &0 & 43 &  0&  0& 40&  3&  0 & 27 &16 & 0] \\ \hline 
middle& 14 &  0 & 1 & 13&  2&  0& 12&  3&  0&9&  5 & 1 \\ \hline 
upper & 6&  0&  2& 6&  0&  2 & 7&  1&  0 & 4 & 2  &2 \\ \hline  
\multicolumn{4}{|p{1.2in}||}{\centering Accuracy = 0.68} &
\multicolumn{3}{|p{0.9in}||}{\centering  Accuracy = 0.71} &
\multicolumn{3}{|p{0.9in}||}{\centering  Accuracy = 0.65} &
\multicolumn{3}{|p{0.9in}|}{\centering  Accuracy =0.52} \\ \hline
\multicolumn{13}{|p{3.9in}|}{\centering {\bf  Recall (R), Precision (P), F1-Score (F1) Evaluation  } } \\ \hline 
\hline 
& R & P & F-1 & R & P & F-1& R & P & F-1& R & P & F-1 \\ \hline
lower& 0.68   &   1.00  &     0.81 & 0.69  &    1.00   &   0.82  &  0.68 &     0.93 &     0.78 &  0.68  &    0.63   &   0.65    \\ \hline 
middle & 0.00  &    0.00  &    0.00 &  1.00  &    0.13  &    0.24&  0.43  &    0.20 &     0.27 &  0.22&      0.33  &    0.26 \\ \hline 
top & 0.67   &   0.25   &   0.36  &  1.00   &   0.25  &   0.40 & 0.00    &  0.00 &     0.00& 0.67 &     0.25  &    0.36 \\ \hline 
\multicolumn{13}{|p{3.9in}|}{\centering {\bf HealthWatcher(Size: 135 classes or interfaces) Confusion Matrix}} \\ \hline 
lower & 47 &  1&  9 & 49&  4&  4 &  41 & 8 & 8 &28& 16& 13 \\ \hline 
middle & 20 &  5& 12 & 15& 20&  2 & 7& 28&  2 &6& 30 & 1  \\ \hline 
upper & 5&  1& 35& 7&  0& 34& 6&  6& 29 &3 & 9& 29 \\ \hline 
\multicolumn{4}{|p{1.2in}||}{\centering Accuracy = 0.64} &
\multicolumn{3}{|p{0.9in}||}{\centering  Accuracy = 0.76} &
\multicolumn{3}{|p{0.9in}||}{\centering  Accuracy = 0.72} &
\multicolumn{3}{|p{0.9in}|}{\centering  Accuracy = 0.63} \\ \hline  \hline  \multicolumn{13}{|p{3.9in}|}{\centering {\bf  Recall (R), Precision (P), F1-Score (F1) Evaluation  } } \\ \hline \hline
& R & P & F-1 & R & P & F-1& R & P & F-1& R & P & F-1 \\ \hline
lower &  0.65  &    0.82 &     0.73  &  0.69  &    0.86 &     0.77 & 0.76  &    0.72  &    0.74 &  0.76&      0.49 &     0.60  \\ \hline
middle & 0.71   &   0.14   &   0.23 &  0.83  &    0.54   &   0.66 & 0.67  &    0.76  &   0.71 & 0.55&      0.81  &    0.65  \\ \hline 
upper &0.62    &  0.85  &    0.72   &   0.85  &    0.83  &    0.84 &  0.74  &    0.71    &  0.72 & 0.66&      0.66 &     0.66 \\ \hline 
\multicolumn{13}{|p{3.9in}|}{\centering {\bf Test Architecture System (Size = 16 Classes) Confusion Matrix}} \\ \hline 
lower & 5 & 2 &0 & 4 & 3 & 0 & 5 & 2 & 0 & 5& 2 &0 \\ \hline 
middle & 1 & 4 &0 & 0 & 5 &0 & 1 & 4 &0 & 1& 4 &0 \\ \hline 
upper & 1 & 0 &3 & 0 & 1 & 3 & 4 &0 &0 & 1& 0 &3 \\ \hline 
\multicolumn{4}{|p{1.2in}||}{\centering Accuracy = 0.75} &
\multicolumn{3}{|p{0.9in}||}{\centering  Accuracy = 0.75} &
\multicolumn{3}{|p{0.9in}||}{\centering  Accuracy = 0.56} &
\multicolumn{3}{|p{0.9in}|}{\centering  Accuracy =0.75} \\ \hline 
\multicolumn{13}{|p{3.9in}|}{\centering {\bf  Recall (R), Precision (P), F1-Score (F1) Evaluation  } } \\ \hline 
& R & P & F-1 & R & P & F-1 & R & P & F-1 & R & P & F-1  \\ \hline 
 lower&       0.71&      0.71    &   0.71 &  1.00&      0.57&      0.73&  0.50 &     0.71   &   0.59 &  0.71   &   0.71  &    0.71  \\ \hline 
 middle  &     0.67  &    0.80&      0.73  &  0.56 &     1.00 &     0.71 &  0.67 &     0.80 &     0.73 &    0.67    &  0.80    &  0.73 \\ \hline    
 upper  &     1.00 &     0.75 &     0.86 & 1.00 &     0.75 &     0.86 & 0.00  &    0.00 &     0.00& 1.00  &    0.75  &    0.86 \\ \hline 
\end{tabular}
\end{center}
\end{table*}

\section{Evaluation}
\subsection{Test cases} The following software systems are used to evaluate the performance of the architecture recovery approach developed in this paper. It includes:
\begin{enumerate}
\item {\bf Test Architecture system}: A  small-scale test architecture system, as shown in Figure \ref{tc}, has been specially designed to test the approach. It is a simulated architecture test case.  It includes   16 classes without the implementation of any functionalities. It includes only dependencies among classes, as shown in  Figure \ref{tc}. The classes named as {\em SEC, TX, SERI} in the figure represent crosscutting concerns.
\item {\bf ConStore:} ConStore is a small scale Java-based library designed to manage concept networks.
The concept network  is a mechanism  used to represent the meta-model of an application domain consisting of concepts and connections between them. The ConStore is a framework for detailing out the concepts and creating a domain model for a given application.  It provides services   to store, navigate and retrieve the concept network\cite{constore}.
\item {\bf HealthWatcher:} The HealthWatcher is a web-based application   providing healthcare-related services\cite{greenwood2007impact}. This
application provides services to users to communicate health-related issues. Users can register, update and query
their health-related complaints and problems. The application follows a client-server, layered architecture style.
\end{enumerate}
All these applications are selected as test cases because the layering of the program elements was known in advance.  
\subsection{Results and Evaluation} The performance of classification models is typically evaluated against measures such as accuracy, precision, recall, and F1-Score \cite{goutte2005probabilistic}. These metrics are derived from a confusion matrix which compares the count of actual class labels for the observations in a given data set and the class labels as predicted by a classification model. Four different metrics are derived by comparing true labels with the predicted labels. These are accuracy, recall, precision and F1-score. Table \ref{acc} shows the performance analysis against these metrics.  The table compares the  performance of algorithmic-centric approach and data-driven approach for all the test cases.
\subsubsection{Accuracy Analysis}
The accuracy is the rate of correction for classification models. Higher  the value of accuracy, better is the model. From the accuracy point of view, one can observe from the Table \ref{acc} that the data-driven approach   performs better as compared to the algorithmic-centric approach. The decision-based classifier preforms better on all the test cases with an average accuracy of  74\%. This is because the performance of algorithmic approach depends on the proper tuning of various configuration parameters. The results shown in Table \ref{acc} are obtained with the values of configuration parameters shown in Table \ref{cpu}. 
\begin{table}[h]
    \begin{center}
    \begin{tabular}{|c|c|c|c|}
    \hline 
      & ConStore&	Healthwatcher&	Test Arch. \\ \hline
 $\delta_{il}$ &	4&	10&	2 \\ \hline
$\delta_{iu}$&	1&	1&	1 \\ \hline
$\delta_{ol}$&	4&	2&	2 \\ \hline
$\delta_{ou}$&	1&	5&	2 \\ \hline
$\delta_{b}$ &	6&	9&	6 \\ \hline
$\delta_{c}$&	0.8&	0.8&	0.6 \\ \hline
$\delta_{e}$&	0.6&	0.5&	0.6 \\ \hline
    \end{tabular}
\end{center}

    \caption{Configuration Parameters used during layer recovery}
    \label{cpu}
\end{table}
The machine learning models automatically learn  and adjust the model parameters   for the better results of accuracy. In case of algorithmic   approach,  the configuration parameter tuning is an iterative process  and need to try different combinations. 

\subsubsection{Recall, Precision, F1-Score Analysis} Recall indicates the proportion of correctly identified true positives while precision is the proportion of correct positive identification. High values of both recall and precision are desired, but it isn't easy to get high values simultaneously for recall and precision. Hence,  F1- score combines recall and precision into one metrics.  From the recall, precision, and F1-score point of view, one can observe from Table \ref{acc} that decision tree-based classifier performs better with the highest F1-score of   0.86 for the upper layer classification of test architecture system.  Recalling class labels with higher precision for {\em middle layer} is a challenging task for all the models described in this paper. This is because of the presence of many not so cleanly encapsulated functionalities in a module at the middle layer and mapping crosscutting concerns to one of the three layers.
\section{Earlier Approaches and Discussion}

Recovering architecture descriptions from the code  has been one of the widely  and continuously explored problem by Software Architecture researchers. This has resulted in a large number of techniques\cite{maqbool2007hierarchical}, survey papers \cite{garcia2013comparative} and books \cite{isazadeh2017source} devoted to the topic. In the context of  these earlier approaches, this section  provides the rationale behind  the implementation decisions taken while developing our approach.
\subsection{Include Dependencies vs Symbolic Dependencies} The recent study reported in \cite{lutellier2017measuring} has recognized that the quality of recovered architecture depends on the  type of dependencies analyzed to recover architecture. The study analyzes the impact of {\em symbolic dependencies} i.e. dependencies at the program identifier level versus {\em include dependencies} i.e. at the level of  importing files or including packages. Further, it emphasizes that symbolic dependencies are more accurate way to recover structural information. The use of include dependencies is error prone owing to the fact that a programmer may include a package  without using it. 

We used {\em include dependencies} in our approach because extracting  and managing  include dependencies are simple as compared to symbolic dependencies. Further, we mitigated the risk of unused packages by excluding these relationship from further analysis. Many programming environments facilitate the removal of unused packages.   One of our objectives was to develop a data-driven approach and cleaning data in this way  is an established practice in the field of data engineering.

\subsection{Unsupervised Clustering vs Supervised classification}

The techniques of unsupervised clustering have been adopted widely to extract high-level architectures through the analysis of dependencies between implementation artefacts \cite{maqbool2007hierarchical}.  These approaches use hierarchical and search-based methods for clustering. These approaches usually take substantial search time to find not so good architectures  \cite{mohammadi2019new}.  One of the advantages of clustering methods is that unlabelled data sets drive these methods. But,  the identified clusters of program elements need to be labelled with appropriate labels. 

Our choice of {\em supervised classification method} is driven by the fact that {\em centrality measures} quantify the structural properties with reference to a node, and relation of the nodes with respect to others.  Processing such quantified values in efficient way  is one of the advantages of many of supervised classification methods. Further, assigning  program elements with layering labels is not an issue if such information is available   from the  previous version of software, which is the case for many re-engineering and modernization projects.   In  the absence of such labelled data set, the approach presented  in the paper can still be adopted in two stages. In the first stage, a tentative layer labelling can be done through algorithmic approach followed by the labelling through supervised classification method. 

The architecture descriptions  extracted by the approach  can be viewed as multiple ways of decomposing a system   rather than a single ground truth architecture which is often difficult to agree upon and laborious to discover \cite{garcia2013comparative}. One of the architecture   out of these extracted architectures can be selected by assessing these architectures for the properties such as minimal layer of violation \cite{sarkar2009discovery,sarkar2010architecture} or satisfaction of a particular quality attribute\cite{isazadeh2017source} or any other project specific criteria. 

\subsection{Choice of Number of Layers}
We described the working of the approach by assuming a three-layer decomposition, But this is not a strict restriction. The algorithmic centric method can be adapted by redesigning rules for additional layers while supervised classification method can be adjusted by relabelling program element with a number of layers considered.

\section{Conclusion}

The paper presents an approach to recover high-level architecture from system implementations. The main highlights of the approach presented in the paper include: (i) The approach uses the centrality measures from the field of Social Network Analysis to quantify the structural properties of an implementation.  (ii) The dependency graph formed by including programming units(i.e. classes in Java) is treated as a network, and centrality measures are applied to extract structural properties. (iii) The paper treats a {\em layer} as a coarsely granular abstraction encapsulating system functionalities. Then paper maps a group programming elements sharing common structural properties manifested through centrality measures to a layer.  (iv) Paper describes two mapping methods for this purpose called algorithmic centric and data-driven.  (v) Overall data-driven methods perform better compared to the algorithmic centric method to map a program element to a layer.

Paper makes particular assumption such as availability of Java-based system implementation; a system is decomposed into three layers and availability of pre-labelled data set for supervised classification. These are the assumption made to simplify the demonstration of the approach and its realization. Hence, these assumptions do not make the approach a restrictive one. However, these assumptions can be relaxed, and the approach is flexible enough to extend.

Exploring the impact of fusing structural properties along with some semantic features such as a dominant concern addressed by a programming element would be an exciting exercise for future exploration.

\bibliographystyle{plain} 
\bibliography{paper}
\end{document}